\documentclass[iop,twocolappendix,appendixfloats]{emulateapj}
\usepackage{apjfonts,amsmath}
\usepackage{color}

\newcommand\beq{\begin{equation}}
\newcommand\eeq{\end{equation}}

\shorttitle{On Detecting Assembly Bias}
\shortauthors{Lin et al.}

\begin{document}

\title{On detecting halo assembly bias with galaxy populations}

\author{
Yen-Ting Lin\altaffilmark{1},
Rachel Mandelbaum\altaffilmark{2},
Yun-Hsin Huang\altaffilmark{3,4},
Hung-Jin Huang\altaffilmark{1,2,5},
Neal Dalal\altaffilmark{6,7,8},
Benedikt Diemer\altaffilmark{3,9},
Hung-Yu Jian\altaffilmark{1,5},
and Andrey Kravtsov\altaffilmark{3}
}

\altaffiltext{1}{Institute of Astronomy and Astrophysics, Academia Sinica, Taipei 10617, Taiwan; ytl@asiaa.sinica.edu.tw}
\altaffiltext{2}{McWilliams Center for Cosmology,  Department of Physics, Carnegie Mellon University, Pittsburgh, PA 15213}
\altaffiltext{3}{Department of Astronomy and Astrophysics, The University of Chicago, Chicago, IL 60637}
\altaffiltext{4}{Department of Astronomy,  University of Arizona, Tucson, AZ 85721}
\altaffiltext{5}{Institute of Astrophysics, National Taiwan University, Taipei 10617, Taiwan}
\altaffiltext{6}{Department of Astronomy, University of Illinois at Urbana-Champaign, Urbana, IL 61801}
\altaffiltext{7}{Kavli IPMU (WPI), UTIAS, The University of Tokyo, Kashiwa, Chiba 277-8583, Japan}
\altaffiltext{8}{Department of Chemistry and Physics, University of Kwa-Zulu Natal, University Road, Westville, KZN, South Africa}
\altaffiltext{9}{Harvard-Smithsonian Center for Astrophysics, Cambridge, MA 02138}

%%%%%%%%%%%%%%%%%%%%%%%%%%%%%%%%%%%%%%%%%%%%%%
%%%%%%%%%%%%%%%%%%%%%%%%%%%%%%%%%%%%%%%%%%%%%%
\begin{abstract}

The fact that the clustering of dark matter halos depends not only on their mass, but also the formation epoch, is a prominent, albeit subtle, feature of the cold dark matter structure formation theory, and is known as assembly bias.  
At low mass scales ($\sim 10^{12}\,h^{-1}M_\odot$), early-forming halos are predicted to be more strongly clustered than the late-forming ones.
In this study we aim to robustly detect the signature of assembly bias observationally, making use of formation time indicators of central galaxies in low mass halos as a proxy for the halo formation history.  Weak gravitational lensing is employed to ensure our early- and late-forming halo samples have similar masses, and are free of contamination of satellites from more massive halos.  For the two formation time indicators used (resolved star formation history and current specific star formation rate), we do not find convincing evidence of assembly bias.  
For a pair of early- and late-forming galaxy samples with mean mass $M_{200c} \approx 9\times 10^{11}\,h^{-1}M_\odot$, the relative bias is $1.00\pm 0.12$.
We attribute the lack of detection to the possibilities that either the current measurements of these indicators are too noisy, or they do not correlate well with the halo formation history.
Alternative proxies for the halo formation history that should perform better are suggested for future studies.

\end{abstract}

\keywords{
cosmology: large-scale structure of Universe, galaxies: formation, galaxies: haloes
}

%%%%%%%%%%%%%%%%%%%%%%%%%%%%%%%%%%%%%%%%%%%%%%%%%%%%%%%%%%%%%%%%%%%%%%
\section{Introduction}
\label{sec:intro}
%%%%%%%%%%%%%%%%%%%%%%%%%%%%%%%%%%%%%%%%%%%%%%%%%%%%%%%%%%%%%%%%%%%%%%

In the cold dark matter (CDM) structure formation paradigm, the spatial distribution and internal structure of dark matter halos depend not only on their mass, but also the formation time, an effect known as  assembly bias
\citep{gao05,croton07}.  For halos of mass below or close to the nonlinear mass scale $M_{\rm nl}$, those that form earlier would cluster more strongly and be more concentrated, while those that form later would be less clustered and concentrated \citep{gao05,zhu06}.  
For much more massive halos, the situation reverses and the later forming ones actually cluster more strongly, although the effect is weaker compared to the low mass case\footnote{Whether the sign change occurs depends on the definition of halo formation time \citep[][see also Section~\ref{sec:sims}]{li08}.} \citep{jing07,wang07,zentner07,dalal08}.

As the halo assembly bias is a very distinct feature of the CDM theory of structure formation, 
several groups have looked for its observational evidence.  
\citet{yang06} are the first to claim the detection of the effect.  Using a galaxy group catalog that extends to low mass systems with just one or two galaxies,
 they find that for halos of similar masses, those with a central galaxy that has a low  star formation rate (SFR)  cluster more strongly than those with a high SFR central galaxy.
If the SFR of central galaxies correlates well with the formation history of the halos in such a way that lower SFR corresponds to earlier formation epoch, then the observed clustering properties would be a manifestation of assembly bias.  

The way halo mass is estimated in the \citet{yang06} group catalog is in spirit similar\footnote{Abundance-matching methods that are used in practice include additional complexity, such as
   scatter between halo properties and observed galaxy properties.} to the abundance matching technique (e.g., \citealt{kravtsov04,tasitsiomi04,conroy06}), that is, a one-to-one correspondence between the total luminosity (or stellar mass) content of a galactic system and its total mass is assumed \citep{yang05b}.  Therefore, at the low halo mass end the halo mass ranking is equivalent to the luminosity (or stellar mass) ranking of the central galaxies.
It is possible that galaxy formation processes would induce large scatter between the luminosity (or
stellar mass) of the central and the host halo mass, and that the mean relationships between galaxy
and host halo properties  may be different for early- and late-type (or red and blue) galaxies.
Some subsequent studies of assembly bias \citep[e.g.,][]{wang13,lacerna14} also employ the later versions of the group catalog generated by \citet{yang06}, 
so if there are systematic biases in the halo mass estimates due to the assumptions made, then the later studies would be subject to those biases as well (see also Section~\ref{sec:yangmass}).

Two key assumptions in the \citet{yang06} analysis are that the current SFR of central galaxies is a good indicator of its  formation history, and that the formation history of a central is intimately linked to that of the host halo.
\citet{wang13} have examined results from semi-analytic galaxy formation models, and found a correlation between the current specific SFR (sSFR, star formation rate per unit stellar mass) and the formation epoch, which indirectly
supports the former assumption.  Furthermore, in the age-matching framework \citep{hearin13,hearin14,watson15}, which is an extension of the subhalo abundance matching technique and 
is capable of reproducing the color-dependence of galaxy clustering, 
the present-day color or sSFR of galaxies is assumed to directly correspond to the (sub)halo formation time
(however see \citealt{masaki13} for a different point of view).
Therefore, both of these assumptions seem to be well founded in theoretical/phenomenological models of galaxy formation (see also Section~\ref{sec:formation}),
and form the foundation for studying assembly bias with galaxy populations.

Instead of using the sSFR, \citet{lacerna14} consider the luminosity weighted-age of the central galaxies as a proxy of the halo formation epoch.  They find that groups with an older central galaxy cluster more strongly than those with a younger central galaxy, in line with the prediction of assembly bias.

Here we present an observational study of the assembly bias that is distinct from previous investigations in several ways.  Although we also employ the group catalog from Yang and collaborators, and assume that the formation history of halos can be inferred from the properties of the central galaxies, we do not rely on the halo mass estimates given in the catalog.  Instead, we use galaxy-galaxy lensing to ensure our early- and late-forming halo samples are of similar masses.  
Second, in addition to sSFR, we also consider the temporally resolved star formation history (SFH) of central galaxies to estimate the formation time of the host halos.  
Third, as any presence of satellites in a central galaxy sample would inevitably bias the halo mass and the large scale clustering measurements, they have to be removed from the sample.  We have taken care of such contaminants in our analysis.
Fourth, it is difficult to infer the halo mass probability distribution for an observed galaxy sample.  We have devised a way to take into account such uncertainties when we compare our data with theoretical expectations.

It is possible to extend the concept of assembly bias from halos to galaxies, such that for galaxies of the same {\it stellar} mass, those that form earlier would cluster more strongly \citep[e.g.,][]{li06,cooper10,wang13}.  
We shall refer to this phenomenon as {\it galactic} assembly bias, and note that in the limit that the stellar mass of a galaxy is a perfect proxy for its dark matter halo mass, the two types of assembly bias are manifestations of the same phenomenon.  However, both the intrinsic scatter in the central galaxy stellar mass--halo mass relation and the presence of satellite galaxies in the galaxy samples may obscure the interpretation of galactic assembly bias.
In this study we thus focus on 
detecting and quantifying the magnitude of {\it halo} assembly bias from central galaxies of low mass halos.

It is also possible to generalize the assembly bias to refer to dependence of clustering and internal structure on {\it any} halo parameters other than the mass (e.g., \citealt{wechsler06,li08,zentner14}).  Throughout this work we follow the original definition and only investigate the effect of halo formation time on the clustering properties.

Detecting and quantifying the magnitude of assembly bias will have far reaching impact on galaxy formation studies, especially for a class of phenomenological model known as the halo occupation distribution (HOD; e.g., \citealt[][]{berlind02}).  In the standard HOD framework, the halo mass is assumed to be the only governing factor controlling galaxy formation.  Blindly inferring the relationship between galaxies and halos from the galaxy clustering properties without accounting for the effect of assembly bias will lead to biased results, as demonstrated by \citet{zentner14}.

The structure of this paper is as follows. In Section~\ref{sec:overview}, we describe our data, including the galaxy and group catalogs, the measurements of sSFR and SFH, two-point correlation function, the stacked lensing signals, and the simulations we use to derive theoretical expectations.
In Section~\ref{sec:prev} we repeat some of the previous attempts at detecting the assembly bias, showing that most of the claimed signal can actually be attributed to halo mass differences.
Then in Section~\ref{sec:ours} we construct galaxy samples that can be used most robustly (as allowed by the quality of our data) to measure the assembly bias.
We discuss caveats and implications of our analysis and future directions in Section~\ref{sec:disc}, and summarize our results in Section~\ref{sec:conclusion}.

Throughout this paper we adopt a {\it WMAP5} \citep{komatsu09} $\Lambda$CDM cosmological model,
where $\Omega_m=0.26$, $\Omega_\Lambda=0.74$, $H_0=100h~{\rm km\,s^{-1}\,Mpc^{-1}}$ with $h=0.71$,
 $\sigma_8=0.8$ and $M_{\rm nl}=2.7\times 10^{12}\,h^{-1}M_\odot$ at $z=0$.
Unless otherwise noted, the halo mass definition we adopt is $M_{200c}$, the mass enclosed in $r_{200c}$, within which the mean overdensity is 200 times the critical density of the Universe $\rho_{\mathrm crit}$
at the redshift of the halo.
Another popular mass definition, $M_{200b}$, defined analogously to $M_{200c}$ but with respect to the mean density of the Universe, will be used in parts of Section~\ref{sec:sfh}.

%%%%%%%%%%%%%%%%%%%%%%%%%%%%%%%%%%%%%%%%%%%%%%%%%%%%%%%%%%%%%%%%%%%%%%
\section{Key Elements of Analysis}
\label{sec:overview}
%%%%%%%%%%%%%%%%%%%%%%%%%%%%%%%%%%%%%%%%%%%%%%%%%%%%%%%%%%%%%%%%%%%%%%

%%%%%%%%%%%%%%%%%%%%%%%%%%%%%%%%%%%%%%%%%%%
\subsection{Catalog of Central Galaxies}
\label{sec:cat}
%%%%%%%%%%%%%%%%%%%%%%%%%%%%%%%%%%%%%%%%%%%

Similar to previous observational studies of assembly bias, we assume that if the formation history of galaxies in a halo could somehow be related to that of the host halo, it is the central galaxy that 
traces best the halo formation history.
Therefore, the starting point of our analysis is a catalog of central galaxies, for which we use the galaxy group catalog of \citet[][hereafter Y07]{yang07}.
The version of the catalog used is based on data release seven (DR7; \citealt{sdssdr7}) of the Sloan Digital Sky Survey \citep[SDSS;][]{york00}.
The Y07 group finding algorithm is essentially an adaptive matched filter applied to the spectroscopic sample of SDSS.  Its unique features include (1) the identification of groups of all richness, ranging from massive clusters down to systems with just one member, (2) designation of central and satellite galaxies among the group members, (3) assignment of total mass of the groups via an abundance matching method (see Section~\ref{sec:intro}; in this study we adopt the mass based on the luminosity content-ranking).
The DR7 group catalog contains 374,052 systems\footnote{Specifically, we use their sample III.  Note that there are 472,416 groups in the DR7 catalog, but we only use those that have a halo mass assignment {\it and} a spectrum measured for the central galaxy.} over 7,748 deg$^2$ out to $z=0.2$ \citep{wang14}.

After selecting the galaxies identified as centrals from the Y07 group catalog, we seek  measurements of sSFR and SFH from public data sets.  For the former we adopt the measurements from the DR7 version of the MPA/JHU value-added galaxy catalog \citep{brinchmann04}, which are derived from emission lines as well as the strength of the 4000\,\AA\  break.

For the resolved SFH, we utilize the results of the VESPA algorithm \citep{tojeiro09}, which are available for query in the WFCAM science archive.
In short, VESPA considers combinations of bursts of star formation of different metallicity, and chooses the one that best fits the observed SDSS spectrum.  For each galaxy, it provides the stellar mass formed in 16 temporal (age) bins, where the first bin spans the period of 9 to 14 Gyr before the redshift of the galaxy (and other bins are closer to the present epoch).  
In addition to using two of the popular stellar population synthesis models \citep{bruzual03,maraston05} to model the spectrophotometric evolution of the galaxies, two different dust extinction models are also considered in VESPA.
Therefore, for each galaxy, four SFHs are provided \citep{tojeiro09}.
For simplicity, we adopt the SFHs based on the \citet{bruzual03} model, and demand the consistency between the SFHs derived from the two dust models (that is, a galaxy should be classified as early- or late-forming in both models).

In the following, unless otherwise stated, an early-forming galaxy is defined to have formed 50\% of
its final stellar mass in the first temporal bin, or roughly by $z=1.9$ for a galaxy observed at
$z=0.1$, which is close to the mean redshift of the samples we use in Section~\ref{sec:ours}.  In contrast, a late-forming galaxy would have formed 50\% of its final mass after the first temporal bin (or at $z<1.9$ if it is observed at $z=0.1$).

%%%%%%%%%%%%%%%%%%%%%%%%%%%%%%%%%%%%%%%%%%%
\subsection{Correlation Function Measurements}
%%%%%%%%%%%%%%%%%%%%%%%%%%%%%%%%%%%%%%%%%%%

To measure the large-scale bias of the galaxy samples, we first calculate the redshift-space two-point auto-correlation function via the standard estimator \citep{landy93}
\begin{equation}
\xi_s = \frac{\langle DD \rangle  -2 \langle DR \rangle + \langle RR \rangle   }{\langle RR \rangle},
\end{equation}
where $\langle DD \rangle$, $\langle DR \rangle$, and $\langle RR \rangle$ are the normalized numbers of data-data, data-random, and random-random pairs in a given separation bin, respectively.
In practice, to bypass complications due to redshift space distortions, we measure $\xi_s$ in two dimensions, both perpendicular to and along the line of sight (denoted as $r_p$- and $\pi$-directions, respectively), then compute the projected correlation function by integrating $\xi_s$ over the $\pi$ direction
\begin{equation}
w_p(r_p) = 2 \int_0^{\pi_{\rm up}} d\pi \xi_s(r_p, \pi).
\end{equation}
Following common practice, the integration upper limit is chosen to be $\pi_{\rm up}=60
\,h^{-1}$\,Mpc, so that the results are not affected by redshift space distortion below $r_p\sim
30\,h^{-1}$Mpc \citep[e.g.,][]{padmanabhan07b}.
As will be described below, our galaxy samples have sizes of several tens of thousands.
We therefore use one million random points over the DR7 footprint, generated with the mask provided by the NYU value-added galaxy catalog \citep{blanton05}.
The covariance matrix of the correlation function is calculated using the jackknife resampling method.  We have divided the DR7 footprint into 100 equal-area subregions, and have constructed the jackknife samples by omitting each of the subregions in turn.  The covariance matrix is then estimated as
\begin{equation}
{\rm C}(w_{p,i}, w_{p,j}) = \frac{N-1}{N} \sum_{\ell=1}^{\ell=N} (w_{p,i}^\ell-\overline{w_{p,i}}) (w_{p,j}^\ell-\overline{w_{p,j}}),
\end{equation}
where $N=100$, 
$\ell$ is the index of the jackknife samples, 
$i,j$ are indices of $r_p$ bins, and $\overline{w_{p,i}}$ is the mean from all the jackknife samples.
Note that $N$ is much larger than the number of radial bins in our correlation function
measurements (see, e.g., appendix D of \citealt{hirata04}, which includes a method of
simulating the impact of noise on the jackknife-based $\chi^2$ that we use here to estimate
$p$-values), and the size of 
the jackknife subregions is much larger than the maximum size of the 
clustering used for the analysis ($\lesssim 35\,h^{-1}$Mpc), so that the samples nearly satisfy the
i.i.d.\ (independent and identically-distributed) requirement.
These conditions validate the use of jackknife resampling.

%%%%%%%%%%%%%%%%%%%%%%%%%%%%%%%%%%%%%%%%%%%
\subsection{Galaxy-Galaxy Lensing Measurements}
\label{sec:lensing}
%%%%%%%%%%%%%%%%%%%%%%%%%%%%%%%%%%%%%%%%%%%

Galaxy-galaxy weak lensing, the coherent tangential shear of background galaxies due to foreground
lens galaxies, provides a simple way to probe the
connection between the lens galaxies and matter via their
cross-correlation function $\xi_{gm}$.  This cross-correlation can be related to the
projected surface density\footnote{Here we neglect the very broad radial window function.} via
\begin{equation}\label{E:sigmar}
\Sigma(r_p) = \overline{\rho} \int \left[1+\xi_{gm}\left(\sqrt{r_p^2 + \pi^2}\right)\right] d\pi.
\end{equation}
The surface density is then related to the observable quantity for lensing, 
\begin{equation}\label{E:ds}
\Delta\Sigma(r_p) = \gamma_t(r_p) \Sigma_c= \overline{\Sigma}(<r_p) - \Sigma(r_p).
\end{equation}
This observable quantity can
be expressed as the product of two factors, a tangential shear
$\gamma_t$ and a geometric factor
\begin{equation}\label{E:sigmacrit}
\Sigma_c = \frac{c^2}{4\pi G} \frac{D_S}{D_L D_{LS}(1+z_L)^2}
\end{equation}
where $D_L$ and $D_S$ are angular diameter distances to the lens and
source, $D_{LS}$ is the angular diameter distance between the lens
and source, and the factor of $(1+z_L)^{-2}$ arises due to our use of
comoving coordinates.

As sources to measure the galaxy-galaxy lensing signal, we use a 
catalog \citep{reyes12}
of 1.2 background galaxies per arcmin$^2$ with weak lensing shears estimated using 
the re-Gaussianization method \citep{hirata03}
and photometric redshifts from Zurich Extragalactic Bayesian Redshift
Analyzer \citep[ZEBRA,][]{feldmann06}.   The catalog is 
characterized in detail in several papers 
\citep[see ][]{reyes12,mandelbaum12,nakajima12,mandelbaum13}.

The lensing measurement begins with identification of background galaxies around each lens (with photometric redshift 
larger than the lens spectroscopic redshift).  Inverse variance weights are assigned to each 
lens-source pair, including both shape noise and measurement error terms in the variance:
\begin{equation}
w_{ls} = \frac{1}{\Sigma_{\rm crit}^{2}(\sigma_e^2 + \sigma_{\rm{SN}}^2)},
\label{eq:wls}
\end{equation}
where $\sigma_e^2$ is the shape measurement error due to pixel
  noise, and $\sigma_{\rm{SN}}^2$ is the root-mean-square intrinsic ellipticity (both quantities are per component, rather than
  total; the latter is fixed to $0.365$ following \citealt{reyes12}). 
Use of photometric redshifts, which have nonzero bias and significant scatter, 
gives rise to a bias in the signals that can be easily corrected using
the method from \cite{nakajima12}.  This bias is a function of lens redshift, 
and is calculated including all weight factors for each lens sample taking into 
account its redshift distribution.  For typical low redshift lens samples in this work, the 
bias is of order 1 per cent, far below the statistical errors.

$\Delta\Sigma$ in each radial  bin can be computed via
a summation over lens-source pairs ``$ls$'' and random lens-source pairs
``$rs$'':
\begin{equation}
\Delta\Sigma(r_p) = \frac{\sum_{ls} w_{ls} e_t^{(ls)} 
\Sigma_{{\rm crit}}(z_l,z_s)}{2 {\cal
    R}\sum_{rs} w_{rs}},
\end{equation}
where $r_p$ is the comoving projected radius from the lens, $e_t$ is the tangential ellipticity component of source
galaxy with respect to the lens position, 
$\mathcal{R} \approx (1 - e_{\rm rms}^{2})$ is the shear responsivity \citep{bernstein02} that converts from the ensemble average distortion
to shear, and $e_{\rm rms}$ is the root-mean-square distortion per component.
The division by $\sum w_{rs}$ accounts for the fact
that some of our ``sources'' are physically associated with the lens,
and therefore not lensed by it \citep[see, e.g.,][]{sheldon04}.  Finally, we subtract off a similar
signal measured around random points with the same area coverage and redshift distribution as the lenses, to subtract off any coherent
systematic shear contributions \citep{mandelbaum05}; this
signal is statistically consistent with zero for all scales used in
this work.

To calculate the error bars, which are dominated by shape noise, 
we use the jackknife resampling method. The maximum
scale used for the fits in the lensing analysis is 
$1 h^{-1}$Mpc, which for a typical lens redshift is far below the 
typical size of each jackknife resampled region.  Thus, the
jackknife method is a reasonable approach to getting the covariance matrix for the projected
mass profile. 

All of the projected mass around lens galaxies contributes to the galaxy-galaxy lensing signal.
This includes contributions from the host dark matter halo in which the lens galaxy resides
(``1-halo term''), and from other dark matter halos (``two-halo term'') that are part of large-scale
structure associated with the lens.  For central galaxies, the 1-halo term simply corresponds to the
$\Delta\Sigma$ for the host dark matter halo.  For satellite galaxies, there are two contributions
to the 1-halo term: on small scales, a contribution from the satellite subhalo, and on larger scales
($0.3$--$2h^{-1}$Mpc) a contribution from the host halo itself.  See, e.g., \cite{mandelbaum05b} for
illustrations of these contributions to the galaxy-galaxy lensing signal.  The distinctive shape of
the satellite contribution in the lensing signal will be a diagnostic of possible contamination of
our ``central galaxy'' sample by satellites in this work.

We fit the observed lensing profiles to the prediction for a pure \citet[][hereafter NFW]{navarro97}
profile.  The NFW profile in principle has two parameters: the concentration and the total mass
within some fiducial radius, for which we use a spherical overdensity of $200\rho_{\mathrm crit}$. 
We fix the relationship between concentration $c_{200}$ and mass $M_{200c}$ using a fitting formula from
\citet{diemer15}.

For a given mass (and therefore concentration), the $\Delta\Sigma$ is predicted via direct
integration.  Finally, we fit for the mass via $\chi^2$ minimization using the Levenberg-Marquardt
algorithm.  This fit is carried out twice for each sample.  In the first case, we fit from $0.04$ to
$0.3\,h^{-1}$Mpc, and compare the best-fit profile with the observed signal out to larger scales,
$1\,h^{-1}$Mpc.  
If the sample is contaminated by satellites, this will be evident in a substantial
excess in the observed signal for $r_p>0.3\,h^{-1}$Mpc compared to the theoretical one fit to $r_p<0.3\,h^{-1}$Mpc and then extrapolated to larger scales.
We use this test as a way to identify samples that are not truly central galaxies and that therefore cannot be used for our
analysis (see Section~\ref{sec:prev}).  If we do not see any sign of satellite contamination, then
we redo the fit using $\Delta\Sigma$ from $0.04$ to $1\,h^{-1}$Mpc (as in Section~\ref{sec:ours}).
As shown directly by \citet{mandelbaum05b} using fits to lensing signals from simulated galaxy samples based on $N$-body simulations, the best-fitting mass will lie between the true median and mean halo mass for the sample, and is most easily interpreted for samples with relatively narrow
mass distributions.

%%%%%%%%%%%%%%%%%%%%%%%%%%%%%%%%%%%%%%%%%%%
\subsection{Numerical Simulations}
\label{sec:sims}
%%%%%%%%%%%%%%%%%%%%%%%%%%%%%%%%%%%%%%%%%%%

To compare our measurements of assembly bias against theoretical expectations, we use three sets of cosmological simulations. The first one is a subset of $N$-body simulations presented in \citet[][specifically the L0250, L0500, and L1000 boxes]{diemer14}.
Each of these simulations follows $1024^3$ dark matter particles. Combined with the box sizes of 250, 500, and 1000$\,h^{-1}$Mpc on a side, the resulting particle mass resolutions are $1.1\times 10^9$, $8.7\times 10^9$, and $7\times 10^{10}\,h^{-1}M_\odot$, respectively. The force resolution lengths are 5.8, 14 and 33 comoving $h^{-1}$kpc, respectively. The cosmological parameters used in these simulations ($\Omega_m=1-\Omega_\Lambda=0.27$, $h=0.7$, $\sigma_8=0.82$, $M_{\rm nl}=3.2\times 10^{12}\,h^{-1}M_\odot$ at $z=0$) are close to the ones adopted in the current study. Halos and merger trees were extracted using the Rockstar \citep{behroozi13} and consistent-trees \citep{behroozi13b} codes. 
For more details of the simulations we refer the reader to \citet{diemer14}.

The main quantity we wish to derive from this set of simulation data is the redshift of formation $z_{\rm form}$ for the halos, for which we consider two commonly used definitions: (1) $z_{50}$, the epoch when a halo has acquired 50\% of its final mass, and (2) $z_{\rm mah}$, which is obtained from the mass accretion history of halos, following the prescription of \citet{wechsler06}.  
We fit the 
mass accretion history of halos by the form
\begin{equation}
M(z) \propto \exp(-\alpha z),
\end{equation}
with the Levenberg-Marquardt algorithm for minimization of the
least squares, and define $z_{\rm mah}=2/\alpha-1$.  
Since $z_{\rm mah}$ is determined by the overall merger history, it is less sensitive to individual events in halo assembly than other definitions of $z_{\rm form}$ \citep{wechsler06}.
In Section~\ref{sec:sfh} we will compare the clustering properties of central galaxies selected by their SFH 
with that of halos, selected by either definition of $z_{\rm form}$.

\begin{figure}
%\epsscale{0.9}
\plotone{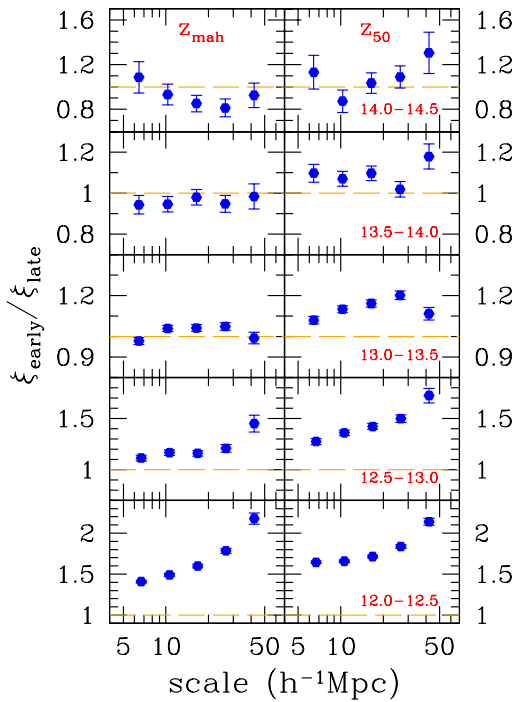}
\vspace{-6mm}
\caption{ 
Ratio of early-to-late-forming halo correlation functions from simulations of \citet{diemer14}.  From bottom to top, each row represents a halo mass bin with width of 0.5 dex, in order of increasing mass (logarithm of halo mass range is indicated on the lower right corner of each panel).  The column on the left is for results obtained with $z_{\rm form}=z_{\rm mah}$.  The right column is obtained with $z_{\rm form}=z_{50}$.
  }
\label{fig:simbias}
\end{figure}

The second set of simulations we use is in the form of a mock galaxy catalog, which is made publicly available by \citet{watson15} and is produced by applying the age-matching model to the Bolshoi simulation \citep{klypin11}.  
The Bolshoi simulation box size is the same as that of the L0250 simulation mentioned above, but with much better mass resolution ($1.35\times 10^8\,h^{-1}M_\odot$).
The cosmology adopted by the Bolshoi simulation is the same as that of \citet{diemer14}.

To create the mock catalog, \citet{watson15} first use the standard abundance matching technique to assign stellar mass to the subhalos 
(by associating stellar mass with maximum circular velocity a subhalo has ever attained).  
Then, for subhalos within a given stellar mass range, a sSFR is further assigned to each subhalo (following the observed distribution of sSFR from galaxies of the same stellar mass range), under the simplifying assumption that 
older subhalos have lower sSFR.  
Assembly bias is therefore {\it maximally built-in} to the properties of the mock galaxies via a
monotonic relationship between age and sSFR.
In the resulting mock catalog that contains 207,950 galaxies, we have information such as the position within the simulation box, designation as central or satellite, the stellar mass and sSFR, and the host halo mass, for each galaxy.
In Section~\ref{sec:ssfr} we will make use of this catalog and compare the clustering properties of central galaxies selected by sSFR with the predictions from the age-matching model.

As our third set of simulations, we make use of the catalog from the semi-analytic model of \citet{guo11}, which is built upon the halo and subhalo merger history extracted from the Millennium Run simulation \citep{springel05}.
The latter follows $2160^3$ dark matter particles in a $(500\,h^{-1}$Mpc$)^3$ box, with the  mass and force resolutions of $8.6\times 10^{8}\,h^{-1}M_\odot$ and 5\,$h^{-1}$kpc, respectively.  The cosmology adopted differs from our default one mainly on $\sigma_8$: $\Omega_m=1-\Omega_\Lambda=0.25$, $h=0.73$, $\sigma_8=0.9$.
The semi-analytic model provides, as a function of redshift, properties such as the total halo mass, stellar mass, broadband color, sSFR, and mean mass weighted stellar age for the galaxies.  We have extracted information for central galaxies  whose present-day halo mass is $M_{200c}\ge 10^{11}\,h^{-1}M_\odot$, and computed $z_{50}$ and $z_{\rm mah}$ for the stellar mass assembly history.  
In Section~\ref{sec:sfh} we will compare the prediction of assembly bias from this model with our observations.

%%%%%%%%%%%%%%%%%%%%%%%%%%%%%%%%%%%%%%%%%%%%%
\subsubsection{ Magnitude of Assembly Bias}
\label{sec:magbias}
%%%%%%%%%%%%%%%%%%%%%%%%%%%%%%%%%%%%%%%%%%%%%

It is informative to examine the expected magnitude of assembly bias as a function of halo mass.
We have used the suite of simulations of \citet{diemer14} to calculate the ratio of real space correlation functions between the early- and late-forming halos, in several halo mass bins.
In each mass bin, we designate halos that have $z_{\rm form}$ that is higher (lower) than the mode of the distribution as early-forming (late-forming).  
For $z_{\rm form}$, we consider both $z_{50}$ and $z_{\rm mah}$.
Note that the way halos are split into early- and late-forming ones here is slightly different from what we will employ when comparing to actual galaxy samples (Section~\ref{sec:ours}).
The results are shown in Fig.~\ref{fig:simbias}.  From bottom to top, each row represents a mass bin in order of increasing halo mass.  The left (right) column is obtained when $z_{\rm form}=z_{\rm mah}$ ($z_{50}$).
For our definition of halo formation time and the halo sample selection, we expect the magnitude of assembly bias to be small -- only appreciable at $\log (M_{200c}/h^{-1}M_\odot)<13$.
This informs our decision to focus on halos around $\log (M_{200c}/h^{-1}M_\odot)\approx 12$ in Sections~\ref{sec:prev} and \ref{sec:ours}.
It is interesting to note that while the amplitude of assembly bias is similar for both definitions of $z_{\rm form}$, the detailed halo mass dependence is somewhat different; we see clearly the sign change at $\log (M_{200c}/h^{-1}M_\odot)\approx 14$ when using $z_{\rm mah}$ (that is, {\it younger} halos are more strongly clustered; see also \citealt{wechsler06,dalal08}), but no evidence for it with $z_{50}$ for all the mass bins we have examined, which is consistent with the results of \citet{li08}.

Finally, from Fig.~\ref{fig:simbias} we see that the magnitude of assembly bias at large scales appears to depend on scales.  This behavior is seen in both the \citet{diemer14} and Millennium simulations.  Such scale dependence of the assembly bias potentially has very significant implications for cosmological constraints using large-scale clustering and
will be investigated in a future publication.

%%%%%%%%%%%%%%%%%%%%%%%%%%%%%%%%%%%%%%%%%%%%%
\subsubsection{ Robustness of Halo Mass Inference from Lensing Measurements }
%%%%%%%%%%%%%%%%%%%%%%%%%%%%%%%%%%%%%%%%%%%%%

A potential concern for our analyses in the following sections is that the halo mass distributions for the early- and late-forming galaxy samples may be so different that the lensing measurements are biased from the true mean values in different ways.
The age-matching mock catalog, together with the particle data from the Bolshoi simulation, provide a way to check this issue.  We have constructed a pair of early- and late-forming mock central galaxy samples, selected by stellar mass (as a proxy of halo mass) and sSFR (as a proxy of formation time, in the sense that low and high sSFR representing early and late formation) such that their projected correlation functions at large scales are similar.
The true mean halo masses for the low and high
sSFR samples are $M_{200c}=1.6\times 10^{12}\,h^{-1}M_\odot$ and $2.3\times 10^{12}\,h^{-1}M_\odot$, respectively.  
The halo mass distributions of the two samples are almost identical except for the peak locations.
We then perform mock lensing observations and determine the best-fit NFW masses to be $1.7\times 10^{12}\,h^{-1}M_\odot$ and $2.3\times 10^{12}\,h^{-1}M_\odot$.
Therefore, 
the ratio of the best-fitting NFW masses from lensing is very similar to the real mean halo masses for these samples.
This exercise alleviates the aforementioned concern.

%%%%%%%%%%%%%%%%%%%%%%%%%%%%%%%%%%%%%%%%%%%%%%%%%%%%%%%%%%%%%%%%%%%%%%
\section{ The Direct Approach: Selection of Halos with Abundance Matching-based Mass}
\label{sec:prev}
%%%%%%%%%%%%%%%%%%%%%%%%%%%%%%%%%%%%%%%%%%%%%%%%%%%%%%%%%%%%%%%%%%%%%%

Before presenting our main results in Section~\ref{sec:ours}, here we first follow the approach adopted in \citet{yang06} and seek  signs of assembly bias.  
However, we add an additional step, which is to use galaxy-galaxy lensing to test the fundamental assumption that the halo masses in the catalog can be used to select early- and late-forming centrals with similar halo masses and without the contamination by satellite galaxies.  These results inform the approach we use in Section~\ref{sec:ours}.

\begin{figure}
\epsscale{0.9}
\plotone{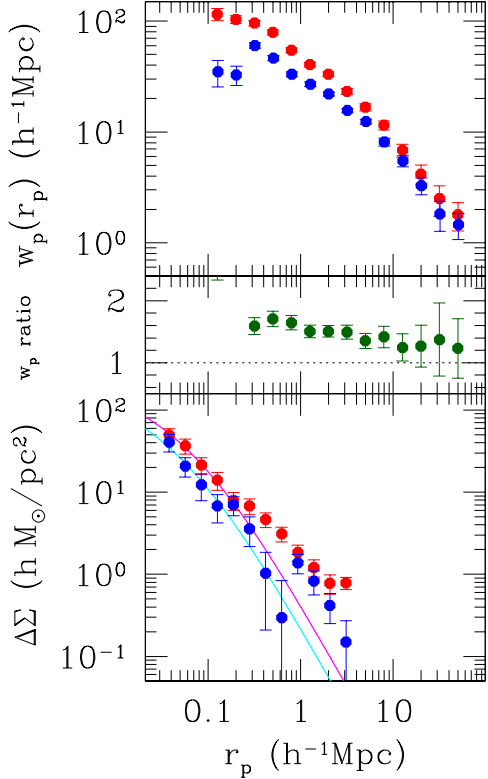}
\vspace{-4mm}
\caption{ 
Measurements of projected correlation function (top panel) and surface mass density contrast (bottom panel) for the central galaxies selected with the Y07 halo mass within the range $\log (M_{200c}/M_\odot)=12.0-12.5$, further separated into low- and high-sSFR samples (red and blue points, respectively).  The middle panel shows the relative bias squared of the two samples: the low-sSFR sample has systematically higher bias, but this is mainly due to the $\sim 1.9$ times difference in halo mass: galaxy-galaxy lensing indicates the two samples have mass $M_{200c}$ of $(9.0^{+1.4}_{-1.2})\times 10^{11}\,h^{-1}M_\odot$ and $(4.6^{+1.0}_{-0.8})\times 10^{11}\,h^{-1}M_\odot$, respectively.  The two curves in the bottom panel represent the best-fit NFW profiles (magenta: low-sSFR; cyan: high-sSFR).
  }
\label{fig:yangssfr}
\end{figure}

We start with the sample of central galaxies with halo mass within the range 
$\log (M_{200c}/M_\odot)=12.0-12.5$,
according to the Y07 catalog. 
We divide the galaxies into high- and low-sSFR samples (containing 75,452 and 61,743 galaxies), with the division at ${\rm sSFR}=10^{-11}$\,yr$^{-1}$.  The projected correlation function and the surface mass density contrast of the samples are shown in the top and bottom panels of Fig.~\ref{fig:yangssfr}, respectively.  In these  panels, the red and blue points represent the low- and high-sSFR samples, respectively.  From the top panel we see that the low-sSFR sample has a systematically higher clustering amplitude.  The ratio of the low-sSFR-to-high-sSFR correlation functions, which represents the relative bias squared of the two samples (at large scales), is shown in the middle panel, and is clearly different from unity (although we note the points are highly correlated).
This is similar to what \citet{yang06} have found (although they have separated the samples using the SFR, not sSFR).
If the two galaxy samples have the same halo mass, then this would represent an observational evidence of assembly bias.

In the bottom panel we show the galaxy-galaxy lensing measurements of the two samples.  They clearly
have different surface density contrasts.  Fitting an NFW model to $\Delta \Sigma$ over $0.04\le
r_p\le 0.3\,h^{-1}$Mpc gives the total halo mass $M_{200c}$ of $(9.0^{+1.4}_{-1.2})\times 10^{11}\,h^{-1}M_\odot$ and $(4.6^{+1.0}_{-0.8})\times 10^{11}\,h^{-1}M_\odot$ for the low-sSFR and high-sSFR samples, respectively.

We note that in the Figure, there are non-negligible contributions from satellite galaxies.  This
can be seen in both the correlation function and galaxy-galaxy lensing measurements.
For the case of the correlation function, in a pure central sample, due to the halo exclusion effect, the signal should flatten at scales below the mean halo radius \citep[e.g.,][]{tinker05}. However, a one-halo term is clearly present in our correlation functions.
For the case of lensing measurements, the ``bump'' from 0.5 to $2\,h^{-1}$Mpc in $\Delta \Sigma$ is also due to the presence of satellites (Section~\ref{sec:lensing}).
Given the mean halo masses of the low-sSFR and high-sSFR samples, we can compute the expected relative bias squared, using the fitting formulae given in \citet{tinker10}.  At the low mass halo regime of our galaxy sample, the bias is a slow varying function of halo mass, resulting in a factor of 1.13 in relative bias squared.  
Given that the measured value is $1.34 \pm 0.19$ (over $5-30\,h^{-1}$Mpc), it seems possible that contamination from satellite galaxies in high mass halos could explain partially the difference in correlation function amplitudes.

\begin{figure}
\epsscale{0.9}
\plotone{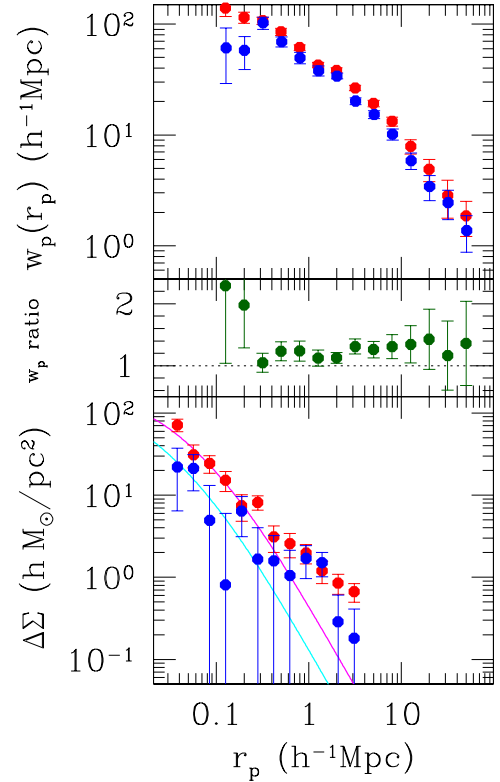}
\vspace{-4mm}
\caption{ 
Similar to Fig.~\ref{fig:yangssfr}, but for central galaxies selected by the resolved SFH from VESPA.  The red and blue points represent the early- and late-forming samples, respectively.
Although the early-forming sample has higher large-scale bias, this is
mainly due to the $\sim 3.4$ times difference in halo mass: galaxy-galaxy lensing indicates the two samples have mass $M_{200c}$ of $(9.7^{+1.9}_{-1.6})\times 10^{11}\,h^{-1}M_\odot$ and $(2.8^{+1.8}_{-1.1})\times 10^{11}\,h^{-1}M_\odot$, respectively.
The two curves in the bottom panel represent the best-fit NFW profiles (magenta: early-forming; cyan: late-forming).
  }
\label{fig:yangsfh}
\end{figure}

We next consider the case where the central galaxy sample is separated by the resolved SFH from VESPA.  We again start with the galaxies within the halo mass range $\log (M_{200c}/M_\odot)=12.0-12.5$ (according to the Y07 catalog).  We define the early-forming galaxy sample to consist of those that have formed 50\% of their final stellar mass in the first temporal bin (containing 63,933 galaxies), while the late-forming galaxies are those that have formed 50\% of the final mass in later bins (56,760 galaxies).
The results are shown in Fig.~\ref{fig:yangsfh}.  The red and blue points now represent the early- and late-forming galaxies, respectively.
Again, although the early-forming sample has a higher bias, this is due to its higher halo mass: the total masses are $(9.7^{+1.9}_{-1.6})\times 10^{11}\,h^{-1}M_\odot$ and $(2.8^{+1.8}_{-1.1})\times 10^{11}\,h^{-1}M_\odot$ for the early- and late-forming samples, respectively.
The presence of satellites is also apparent from both the correlation function and lensing measurements.
The relative bias squared from the halo mass difference alone is 1.25, consistent with the measurement ($1.29 \pm 0.31$).

We therefore conclude that it is likely that either the mean relationship between total luminosity
and halo mass is incorrect for some subsamples of the group catalog, or that the scatter in the Y07 halo
mass estimates is not random but rather correlates with physical properties of the galaxies (such as sSFR and SFH). 
Simply taking the halo mass estimates from the Y07 catalog would lead to false detections of
assembly bias.  We also conclude that some of the central galaxies in the catalog are actually
misidentified satellites.

%%%%%%%%%%%%%%%%%%%%%%%%%%%%%%%%%%%%%%%%%%%%%%%%%%%%%%%%%%%%%%%%%%%%%%
\section{ Our Two-Step Approach: Halo Mass Measurements from Weak Lensing}
\label{sec:ours}
%%%%%%%%%%%%%%%%%%%%%%%%%%%%%%%%%%%%%%%%%%%%%%%%%%%%%%%%%%%%%%%%%%%%%%

Equipped with the experience gained from the exercises in Section~\ref{sec:prev}, we now present our approach to the detection of assembly bias.
Our goal is to construct early- and late-forming central galaxy samples that have similar halo masses.
As in Section~\ref{sec:prev}, we classify galaxies as early- or late-forming via either the resolved SFH from VESPA or the current sSFR.
Below we describe our procedures and results with these two methods in turn.

%%%%%%%%%%%%%%%%%%%%%%%%%%%%%%%%%%%%%%%%%%%
\subsection{Classification by Resolved Star Formation History}
\label{sec:sfh}
%%%%%%%%%%%%%%%%%%%%%%%%%%%%%%%%%%%%%%%%%%%

The most important finding from Section~\ref{sec:prev} is that, after identifying samples with the same Y07 halo mass estimates and splitting by either the SFH or sSFR, the mean halo mass of the resulting early- and late-forming samples would actually be quite different.  This makes such a procedure an invalid way to test for assembly bias.
This then motivates us to develop a two-step approach: 
we can start with two central galaxy samples (denoted as samples 1 and 2) that are selected by certain halo mass proxy, according to which the mean masses $M_1<M_2$.  Referring to the mean halo masses of the resulting early- and late-forming subsamples after splitting the samples by either the SFH or sSFR as $M_{1e}$, $M_{1l}$, $M_{2e}$, and $M_{2l}$ (so that in general $M_{1e}>M_{1l}$ and $M_{2e}>M_{2l}$), which are then {\it measured} by weak lensing, we can achieve $M_{1e} \approx M_{2l}$ (e.g., within $1\sigma$) 
by adjusting the chosen range of mass proxy of sample 2 ($M_2$) relative to that of sample 1 ($M_1$).

It is also found in Section~\ref{sec:prev} that there are non-negligible satellite contamination in the Y07 central galaxy sample.  
As these satellites are typically residing in halos more massive than the ones that host our central galaxies, their large-scale bias reflects that of their hosts, and therefore the inclusion of these satellites would dilute any assembly bias signal we are after.
Below we describe details of the steps that lead to our final samples.

%%%%%%%%%%%%%%%%%%%%%%%%%%%%%%%%%%%%%%%%%%%
\subsubsection{Sample Construction}
\label{sec:sampleconstruction}
%%%%%%%%%%%%%%%%%%%%%%%%%%%%%%%%%%%%%%%%%%%

\begin{figure}
\plotone{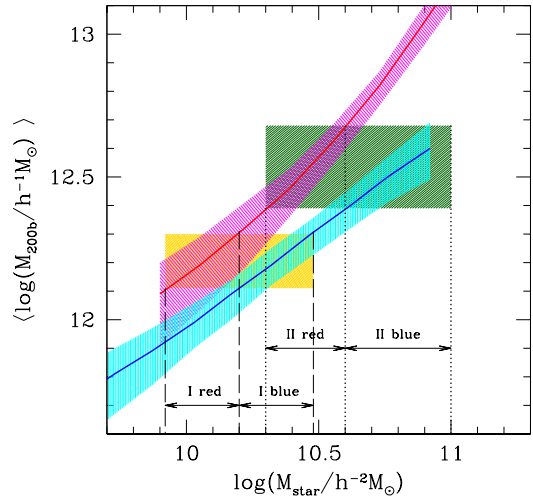}
\vspace{-3mm}
\caption{ 
The red and blue curves represent the central galaxy stellar mass--halo mass relations for the red and blue centrals, respectively, from \citet{more11}.  The shaded regions bounding the curves represent the $1\sigma$ uncertainty around the mean.  
The two horizontal bands represent the first step in our sample construction (i.e., defining the samples 1 and 2).  By selecting red and blue galaxies with stellar mass in the ranges where the horizontal band crosses over the two curves (delineated by the vertical dashed and dotted lines), we obtain galaxy samples that may have similar halo masses.
These samples are further split into early- and late-forming galaxies by the VESPA-based SFH.  Using weak lensing it is found that the late-forming galaxies derived from the samples represented by the green horizontal band and the early-forming galaxies originated from the samples defined by the dark red horizontal band have similar mean halo masses.
We note that the halo mass definition in the M11 relations is $M_{200b}$, instead of $M_{200c}$ as adopted throughout our analysis.  As we use these relations for adjusting the relative masses of galaxy samples, the absolute mass scale (and thus the mass definition) is not important for our purpose.  
  }
\label{fig:surhud}
\end{figure}

Our starting point is a halo mass proxy that can help define samples 1 and 2.  For the SFH-selected samples, we choose to use the central galaxy stellar mass--halo mass relationship for this purpose, and adopt the measurements by \citet[][hereafter M11]{more11}, which are done separately for red and blue centrals (see the solid curves in Fig.~\ref{fig:surhud}), using satellite kinematics.
Using these relations as an approximate guideline, we can select central galaxies in certain stellar mass ranges so that they 
might live in halos of similar masses.  
As suggested by Fig.~\ref{fig:surhud}, we have to combine blue centrals of higher stellar mass with red ones of lower stellar mass\footnote{This Figure provides a way to understand the relative halo masses for the low- and high-sSFR samples studied in Section~\ref{sec:prev}.  Selection by sSFR is equivalent to a color selection.  We also know red galaxies typically have larger stellar mass-to-light ratio than blue galaxies do.  Since the Y07 halo mass estimates are based on the luminosity content of the groups which, at the low mass scale we study, is the same as the luminosity of the central galaxies, this implies the low-sSFR sample would have higher stellar mass than the low-sSFR one; from Fig.~\ref{fig:surhud} we see it is natural for the two samples to have different halo masses.}.
The VESPA-based SFH is then used to classify the galaxies into early- and late-forming ones, for which mean halo masses are measured by galaxy-galaxy lensing.

In implementing the above procedure, we follow M11 and estimate the stellar mass via
\begin{equation}
\begin{split}
\log \Big(  \frac{M_{\rm star}}{h^{-2} M_\odot}  \Big)  = & -0.406 + 1.097 (g-r)  \\ 
        & - 0.4(M_r - 5\log h -4.64)
\end{split}
\end{equation}
(see also \citealt{bell03}).
Here $M_r$ is the SDSS $r$-band absolute magnitude, $(g-r)$ is estimated in the rest-frame; for both of these we use the SDSS Petrosian magnitudes with $k$-corrections taken from the NYU value-added galaxy catalog \citep{blanton05}.  
After examination of our galaxies in the color-stellar mass space, we adopt a red/blue division line that is somewhat
different from that used in M11:
\begin{equation}
(g-r)_{\rm div} = 0.67 + 0.03  \log [ M_{\rm star} / (10^{10} h^{-2} M_\odot ) ]. 
\end{equation}

The next step in our procedure is the removal of satellites. We have found that satellite contamination is more serious in red galaxy samples (i.e., more enhanced 1-halo term in the galaxy clustering signal).  
Thus, in the above process, after a red central sample is selected, we apply a FOF algorithm (see \citealt{jian14} for the description of the code) to identify groups of galaxies, using comoving linking lengths along and perpendicular to the line-of-sight of 1 and 0.08 (in units of the mean galaxy separation)\footnote{
Since we use a flux-limited sample, the mean galaxy separation varies with redshift.  The resulting comoving linking lengths range from 0.18 to $1.92\,h^{-1}$Mpc (perpendicular to the line-of-sight) and from 2.20 to $23.96\,h^{-1}$Mpc (along the line-of-sight).
We note that our choices of linking lengths (100\% and 8\% of the mean galaxy separation along and perpendicular to the line-of-sight) have not been rigorously tested against mock catalogs; rather they are only validated by inspection of the lensing and clustering signals to confirm the absence of a 1-halo (satellite) term.
}, respectively.
We only keep the most massive member in the groups identified.
During this step, we consider not only the red galaxy sample in question, but also other red centrals in the Y07 central catalog, as well as blue centrals in halos with $\log (M_{200b}/h^{-1}M_\odot)>12.5$ (according to the M11 relations).
Then, after combining the blue central sample with the satellite-trimmed red central sample, the FOF algorithm is applied again to further remove remaining satellites.  Only after these steps do we separate the sample into early- and late-forming ones.
About $7-15$\% of galaxies are removed this way.

It should also be noted that not every galaxy can be unambiguously identified as early- or late-forming from the VESPA SFH, as we have opted for purity and demanded consistency between the two dust models (Section~\ref{sec:cat}).  The early- and late-forming sample sizes would therefore be further reduced (by $5-17\%$) compared to the parent samples.

\begin{figure}
\epsscale{0.9}
\plotone{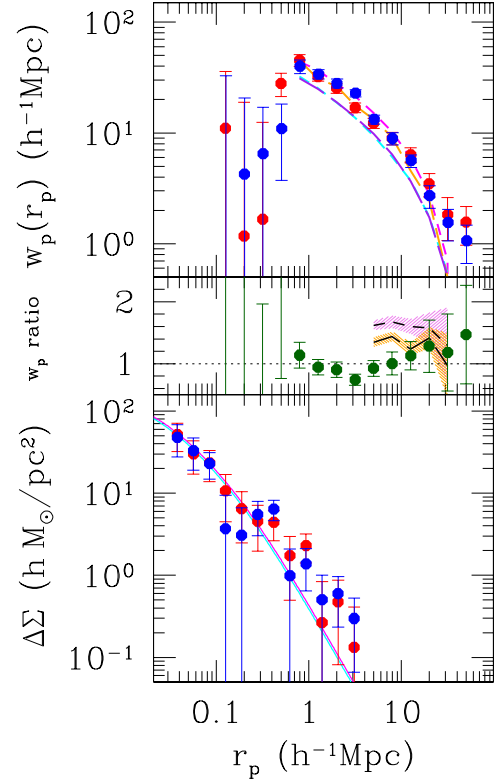}
\vspace{-4mm}
\caption{ 
Measurements of projected correlation function (top panel) and surface mass density contrast (bottom panel) for the central galaxies initially selected with the M11 central-halo relations (see the green and dark red horizontal bands in Fig.~\ref{fig:surhud}), then further separated into early- and late-forming samples using the VESPA-based SFH (red and blue points, respectively).  The points in the middle panel shows the relative bias squared of the two samples.   
The square root of the mean ratio (over $5-35\,h^{-1}$Mpc) is $1.00\pm 0.12$.
Galaxy-galaxy lensing indicates the two samples have mass $M_{200c}$ of $(9.5^{+2.5}_{-2.0})\times 10^{11}\,h^{-1}M_\odot$ and $(8.4^{+2.2}_{-1.8})\times 10^{11}\,h^{-1}M_\odot$, respectively.
The curves in the bottom panel represent the best-fit NFW profiles (magenta: early-forming; cyan: late-forming).
In the top panel, the two short dashed curves show the predictions for the early- and late-forming halos from the L0250 simulation.  The long dashed curves are those from the \citet{guo11} model, based on the Millennium simulation.
The ratios of the two sets of curves (early-to-late) are shown as the curves in the middle panel (short dashed: L0250; long dashed: Millennium), with the shaded regions representing the $1\sigma$ uncertainties from the models. 
These models are found to be inconsistent with the observations (see text for details). 
  }
\label{fig:lmsfh}
\end{figure}

We have thus obtained a pair of early- and late-forming samples 
that we believe are dominated by central galaxies in their halos and have similar mean halo masses.
The way they are constructed is depicted in Fig.~\ref{fig:surhud}.   
In the Figure, two horizontal bands (shaded regions) are shown around $\log (M_{\rm 200b}/h^{-1}M_\odot) \approx 12.4$.  
Each band denotes the initial halo mass range ($M_1$ and $M_2$ in the notation used above) guessed from the central stellar mass-halo mass relations.
The green one represents sample 2 from which a late-forming sample is derived.  
Similarly, the yellow band represents sample 1 from which an early-forming sample is obtained.
More specifically, the log stellar mass ranges for which the red and blue galaxies are selected are $9.9-10.2$ and $10.2-10.45$ ($10.35-10.6$ and $10.52-11.1$) for sample 1 (2).
These ranges are delineated by the vertical dashed and dotted lines.
The resulting early- and late-forming samples contain 18,200 and 26,071 galaxies, respectively.
Using galaxy-galaxy lensing, it is found that these samples have mean masses ($M_{200c}$) that are statistically
consistent: 
$M_{1e}=(9.5^{+2.5}_{-2.0})\times 10^{11}\,h^{-1}M_\odot$ and $M_{2l}=(8.4^{+2.2}_{-1.8})\times 10^{11}\,h^{-1}M_\odot$ (see bottom panel of Fig.~\ref{fig:lmsfh}).  
Below we refer to these as our SFH samples.

In the top and middle panels of Fig.~\ref{fig:lmsfh}, we show the comparison of the correlation functions for these samples.
It is clear that the power at small scales (e.g., $\lesssim 1\,h^{-1}$Mpc) is much reduced compared to Figs.~\ref{fig:yangssfr} and \ref{fig:yangsfh}, due to the removal of satellites with our FOF procedure.
(The lensing signal shown in the bottom panel matches the NFW fit for $r_p\lesssim 2\,h^{-1}$Mpc,
which further supports the conclusion that we do not have an appreciable satellite population in the final galaxy samples.)
At large scales ($r_p \gtrsim 5\,h^{-1}$Mpc), we see in the middle panel the (square of) relative bias scatters around unity (indicated by the horizontal dotted  line), suggesting that there is not a strong difference in the large-scale clustering of the two samples.
Over the scales $5-35\,h^{-1}$Mpc, the square root of the mean ratio is $1.00\pm 0.12$.

%%%%%%%%%%%%%%%%%%%%%%%%%%%%%%%%%%%%%%%%%%%
\subsubsection{Comparison with Theoretical Expectations}
\label{sec:cfmodel}
%%%%%%%%%%%%%%%%%%%%%%%%%%%%%%%%%%%%%%%%%%%

To properly interpret our findings, we first compare the observed relative bias with that obtained from the $z=0$ output of the L0250  simulation (Section~\ref{sec:sims}).  
Note that observationally we can only robustly measure the mean halo mass of our galaxy samples, but not the form of their halo mass {\it distribution}.
To proceed, we thus
 make the assumption that the distribution of halo mass of the observed galaxy samples follows a log-normal form\footnote{The motivation for adopting this form comes from the mock catalog of the age-matching model (Section~\ref{sec:sims}).  For each galaxy in our sample, we assign a counterpart in the mock by matching the stellar mass and sSFR, and find that the resulting halo mass distribution of the matched mock galaxy sample is very close to log-normal, with a standard deviation of $0.2-0.3$.}, 
with parameters $M_{\rm cen}$ and $\sigma_{\log{M}}$ representing the mean and standard deviation of the Gaussian in log space.  To find out the values of  $M_{\rm cen}$ and $\sigma_{\log{M}}$ appropriate for our galaxy samples, we consider a grid of combinations of these parameters, with $\log M_{\rm cen}$ ranging from 11.4 to 12.5 (with an interval of 0.1 dex), and $\sigma_{\log{M}}$ from 0.05 to 0.45 (with an interval of 0.05).
Given an observed galaxy sample, for each combination of $(M_{\rm cen}, \sigma_{\log{M}}$), we predict the corresponding $\Delta \Sigma$ signal and obtain $\chi^2$ by comparing with the observed profile.  We consider all models that satisfy $\chi^2\le \chi_{\rm min}^2+2.3$ as plausible (corresponding to 68\% probability interval for two parameters\footnote{Our conclusion remains unchanged if the criterion for acceptable models is changed to $\chi^2\le \chi_{\rm min}^2+4.61$ (i.e., 90\% probability distribution for two parameters).}), 
where $\chi_{\rm min}^2$ is given by the model with minimum $\chi^2$ on the grid. 
To further split the halo samples into early- and late-forming ones, we proceed as follows.
Given the median redshift $z_{\rm med}$ of an observed galaxy sample, 
we compute the redshift $z_{\rm div}$ corresponding in lookback time 9\,Gyr prior to $z_{\rm med}$, and use it as a proxy for the boundary of the first temporal bin in VESPA.  That is, an early-forming (late-forming) galaxy sample typically has formed 50\% of its final stellar mass before (after) $z_{\rm div}$.
$z_{\rm div}$ for our early- and late-forming galaxy samples are 1.84 and 2.11, respectively.
For each of the models that can represent an observed early-forming (late-forming) galaxy sample, we select halos with $z_{\rm form}>z_{\rm div}$ ($z_{\rm form}<z_{\rm div}$) from the simulation  and compute the correlation function.
Our theoretical expectation is then the average over the correlation functions from all these models.
The uncertainties are estimated 
from jackknife resampling after splitting the simulation box into 125 sub-cubes.
We note that the acceptable models show a degeneracy between $M_{\rm cen}$ and $\sigma_{\log{M}}$ (in the sense that the lower the mass, the higher $\sigma_{\log{M}}$), and have $M_{\rm cen}$ ranging from 11.6 to 12.0, and $\sigma_{\log{M}}$ from 0.05 to 0.45,
and the spread in correlation functions from these models is much smaller (5-6\%) than the mean values.
The fraction of early-forming halos among the halos used in the models is 0.37, fairly close to the observed sample (0.41).

In the top panel of Fig.~\ref{fig:lmsfh}, we show as magenta (cyan) short dashed curve the projected correlation function from the early-forming (late-forming) halo sample. 
Here we have adopted $z_{\rm mah}$ for $z_{\rm form}$, but using $z_{50}$ 
leads to similar conclusions.
The ratio $b^2_{\rm th}(r_p)$ of the early-forming-to-late-forming halo correlation functions from the simulation is shown as the dashed curve in the middle panel of the Figure, with the (pink) shaded region enclosing the $1\sigma$ uncertainties from the model.
Using the covariance matrices built from the ratio of the early- and late-forming projected correlation functions and their associated jackknife samples for both the observations and models ($O$ and $T$, respectively), 
we infer the probability of the observed and theoretical relative bias squared to be consistent with each other from
\begin{equation}
\chi^2 = \sum_{ij}  \bigl(  w_{e}(r_{i})/w_{ l}(r_i) - b^2_{\rm th}(r_i)   \bigr)    D_{ij}^{-1}    \bigl(  w_{e}(r_j)/w_{l}(r_j) - b^2_{\rm th}(r_j)  \bigr),
\label{eq:cov}
\end{equation}
where $w_{e}$ and $w_{l}$ are the observed early- and late-forming projected correlation functions, and $D=O+T$.
For notational simplicity, we have omitted the subscript $p$ in $w_p$ and $r_p$.
For these calculations, we consider only $r_p$ bins in the $5-35\,h^{-1}$Mpc range.
We find that the theory and our observation are inconsistent.  Given 5 degrees of freedom, the probability to yield a $\chi^2$ as large as observed (27.3) if the data and theory were drawn from the same distribution is only  $p=5.0\times 10^{-5}$.
The corresponding $\chi^2$ when $z_{50}$ is adopted is  $64.1$, corresponding to a probability well below
$10^{-10}$.

We next turn to the comparison with the predictions from the \citet{guo11} semi-analytic model.  
Analogous to the approach described above, for a given central galaxy sample, we again start with simulated halo samples permitted by the observed $\Delta \Sigma$ profile, then select early-forming (late-forming) halos as those hosting central galaxies with $z_{\rm form}>z_{\rm div}$ ($z_{\rm form}<z_{\rm div}$).  Here $z_{\rm form}=z_{\rm mah}$ is calculated from the stellar mass assembly history from the \citet{guo11} model.  The uncertainties in the model predictions are also calculated with jackknife resampling of the simulation box.  
The projected correlation functions for the early- and late-forming models are shown as orange and purple long dashed curves in the top panel of Fig.~\ref{fig:lmsfh} respectively, and the ratio of the two is shown as the long dashed curve in the middle panel (the orange shaded region represents the $1\sigma$ uncertainties).
Using Eqn.~\ref{eq:cov}, we find that the probability that the model and data are consistent is $p=0.019$ with $\chi^2=13.5$.
We note, however, that the fraction of early-forming halos among the halos used in the models is only 0.04, far below that of the observed sample.
Given that this model also produces a sSFR distribution that is not consistent with the observed sSFR, we will not further consider it in Section~\ref{sec:ssfr}.

It is possible that at mass scales lower than $M_{\rm nl}$ (as is the case for our samples), the origin of the assembly bias is partially due to the so-called ``backsplash'' halos \citep{dalal08}, those that have been accreted onto massive halos but are on highly elongated orbits and thus would have spent a substantial amount of time outside of the parent halo's virial radius \citep{wetzel14}.
The member galaxies of these backsplash halos would appear old because they have been affected by dense environments of the massive halos, and the bias of these small halos would be that of their massive parent halos.
Our treatment of satellite removal may  exclude some of such halos from our sample, and therefore inadvertently suppress the effect of assembly bias.
Unless there are clever ways to observationally distinguish the backsplash halos from low mass halos that are unrelated to nearby, massive halos, 
it seems to be a challenge to analyses like ours to only
remove satellites that are bounded within massive halos, but not backsplash galaxies.
Perhaps a better (and more feasible) approach is to mimic the effect of satellite removal in simulated data. 
Although a full-blown analysis is beyond the scope of the present paper, in Section~\ref{sec:samples} we will use a simple method to roughly estimate the 
effect of possible exclusion of backsplash halos
to the non-detection of assembly bias.

Ideally, we would like to explore the assembly bias with halos over a wide mass range, similar to what is done in \citet{yang06}.  Unfortunately, at the low mass end, we are limited by the inability of VESPA to yield sufficient number of early-forming galaxies (likely due to the quality of SDSS spectra), thus rendering the lensing measurement too noisy, while at higher mass end, both the facts that the number of central galaxies decreases precipitously, and that most of them are quite old (beyond the temporal resolution offered by VESPA), make it more difficult to apply our method.

%%%%%%%%%%%%%%%%%%%%%%%%%%%%%%%%%%%%%%%%%%%
\subsection{Classification by Specific Star Formation Rate}
\label{sec:ssfr}
%%%%%%%%%%%%%%%%%%%%%%%%%%%%%%%%%%%%%%%%%%%

\begin{figure}
\epsscale{0.9}
\plotone{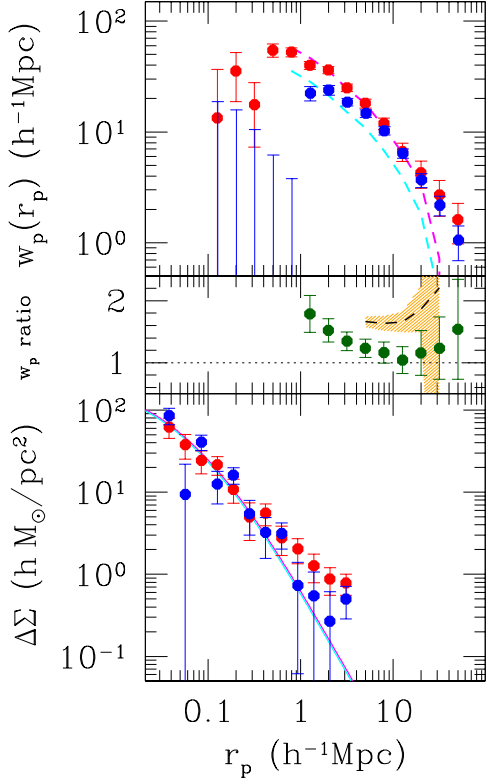}
\vspace{-4mm}
\caption{ 
Measurements of projected correlation function (top panel) and surface mass density contrast (bottom panel) for the central galaxies initially selected with the halo mass estimates given in Y07, then further separated into low- and high-sSFR samples (red and blue points, respectively), with the division at sSFR\,$ =10^{-11.8}$\,yr$^{-1}$.  The middle panel shows the relative bias squared of the two samples.   
The square root of the mean ratio (over $5-35\,h^{-1}$Mpc) is $1.07\pm 0.14$.
Galaxy-galaxy lensing indicates the two samples have mass $M_{200c}$ of $(1.39^{+0.24}_{-0.21})\times 10^{12}\,h^{-1}M_\odot$ and $(1.26^{+0.24}_{-0.20})\times 10^{12}\,h^{-1}M_\odot$, respectively.  
The curves in the bottom panel represent the best-fit NFW profiles (magenta: low-sSFR; cyan: high-sSFR).
In the top panel the two dashed curves show the predictions from the age-matching model \citep{watson15}.  The ratio of the model curves (low-to-high sSFR) is shown in the middle panel.  
After taking into account for the model uncertainties (represented by 
the shaded regions), we find that the model
 is inconsistent with the observations.
}
\label{fig:lmssfr}
\end{figure}

We next follow a procedure similar to that outlined in Section~\ref{sec:sfh} for samples selected by sSFR.  
As we now need to consider high- and low-sSFR centrals separately, operationally it is much more convenient to use the mass estimates from the Y07 catalog, rather than using the M11 relations, in selecting samples 1 and 2.
In constructing the galaxy samples, we again use the FOF algorithm to remove contaminations from satellite galaxies.

We have thus arrived at a pair of low- and high-sSFR samples with similar mean masses.
They are selected by $\log (M_{200c}/M_\odot)=12.0-12.5$ and ${\rm sSFR}<10^{-11.8}\,$yr$^{-1}$, and $\log (M_{200c}/M_\odot)=12.75-13.10$ and ${\rm sSFR}>10^{-11.8}\,$yr$^{-1}$, respectively.
After the FOF step that removes about 10\% of the galaxies, the two samples contain 25,838 and 29,659 galaxies. 
The lensing measurements are shown in the bottom panel of Fig.~\ref{fig:lmssfr}, giving the mean
masses of $M_{200c}=(1.39^{+0.24}_{-0.21})\times 10^{12}\,h^{-1}M_\odot$ and $(1.26^{+0.24}_{-0.20})\times 10^{12}\,h^{-1}M_\odot$. 
The correlation functions and the ratio of low-to-high-sSFR correlation functions are shown in the top and middle panels.  
The square root of the mean ratio (over $5-35\,h^{-1}$Mpc) is $1.07\pm 0.14$.
Given that the data points are all above unity, 
we use a procedure analogous to that described in Section~\ref{sec:sfh} to  examine the consistency between the two samples by calculating 
\begin{equation}
\chi^2 = \sum_{ij}  \bigl(  w_{ls}(r_{i}) - w_{hs}(r_i)  \bigr)    O_{ij}^{-1}    \bigl(  w_{ls}(r_j) - w_{hs}(r_j) \bigr),
\end{equation}
where $w_{ls}$ and $w_{hs}$ are the projected correlation function of the low-sSFR and high-sSFR samples,
and $O$ the covariance matrix built from the difference between $w_{ls}$ and $w_{hs}$ and their associated jackknife samples.
With $\chi^2=5.5$ from 5 degrees of freedom, we find that the two samples have a 36\% probability to be consistent.

In the top and middle panels of the Figure, we compare our measurements with predictions from the age-matching model (\citealt{watson15}; see Section~\ref{sec:sims}).  
Following the procedure outlined in Section~\ref{sec:sfh}, we construct halo samples that can represent the observed galaxy samples by approximating the halo mass distribution as log-normal, and considering halo samples that provide good fits to the observed $\Delta \Sigma$ profile.
We adopt the same division criterion for low- and high-sSFR as in the real data (${\rm sSFR}=10^{-11.8}$\,yr$^{-1}$).
From the curve in the middle panel (representing the ratio of the two model curves shown in the top panel) we see that the age-matching model 
prediction for $b^2_{\rm th}$ is $\sim 1.6$.
We find $(\chi^2,p)=(10.5,0.033)$, indicating the model is inconsistent with the observation.

As the age-matching model is calibrated against a SDSS galaxy sample that is slightly different from the one we use (\citealt{watson15}), it is possible the sSFR distribution in the model does not match perfectly with our data.  After inspecting the cumulative sSFR distributions in both the real data and the model, we find it more appropriate to set the sSFR division to be $10^{-11.4}$\,yr$^{-1}$ in the model for the comparison.  
With this adjustment, the fraction of early-forming halos in the model is 0.41, which is not far from the observed value of 0.47.
The results become $(\chi^2,p)=(20.5,4.0\times 10^{-4})$.
Therefore, our observations do not seem to be compatible with the age-matching model, in which the magnitude of assembly bias is expected to be maximal (Section~\ref{sec:sims}).
Such a result may not be surprising, given that the central galaxy stellar mass--halo mass relationships based on the age-matching model (particularly that for blue galaxies) do not match well with the real data, as indicated by lensing measurements \citep{mandelbaum15}.
%

%%%%%%%%%%%%%%%%%%%%%%%%%%%%%%%%%%%%%%%%%%%%%%%%%%%%%%%%%%%%%%%%%%%%%%
\section{Discussion}
\label{sec:disc}
%%%%%%%%%%%%%%%%%%%%%%%%%%%%%%%%%%%%%%%%%%%%%%%%%%%%%%%%%%%%%%%%%%%%%%

Although we have 
attempted to make
the comparisons between observations and theoretical predictions as fair as possible, there are caveats in our analysis that we need to point out, which may be applicable for other studies of assembly bias as well.  These can be broadly categorized into two themes: (1) definition of formation epoch, and (2) treatment of galaxy and halo samples.  We discuss these aspects in turn in Sections~\ref{sec:formation} and \ref{sec:samples}.
After addressing these potential concerns, 
in Section~\ref{sec:yangmass} we comment on the halo mass estimates of Y07 and M11, and
in Section~\ref{sec:implications} we consider the implications of our findings.

%%%%%%%%%%%%%%%%%%%%%%%%%%%%%%%%%%%%%%%%%%%
\subsection{Definition of Formation Time}
\label{sec:formation}
%%%%%%%%%%%%%%%%%%%%%%%%%%%%%%%%%%%%%%%%%%%

A proper definition of both halo and galaxy formation time is critical, as it affects the expected amplitude of assembly bias as a function of halo mass, and more importantly, it determines whether one can faithfully link the observed galaxy population to the underlying dark matter halos.
As we have seen before, the relative bias between early- and late-forming halos is a function of halo mass, and the detailed mass dependence actually depends on whether $z_{50}$ or $z_{\rm mah}$ is adopted for $z_{\rm form}$ (Fig.~\ref{fig:simbias}).  This in turn affects any comparison with observations, especially the inference of the statistical significance (c.f.~Section~\ref{sec:sfh}).

How can we best link the observed properties of galaxies to the formation history of the host dark matter halos?
Various groups have adopted different ways to define the formation time for the central galaxies, including the current sSFR (or equivalently, broad band optical color) and luminosity-weighted mean age of the stellar populations (Section~\ref{sec:intro}).
One of the new aspects of our study is the use of resolved SFH from the VESPA algorithm.
To gain insight into the answer to the above question, 
we make use of the results from the semi-analytic model of \citet{guo11}.
We have extracted information from $\sim 157,000$ central galaxies  whose present-day halo mass is $M_{200c}= (1-2)\times 10^{12}\,h^{-1}M_\odot$, 
and computed $z_{50}$ and $z_{\rm mah}$ for both the total mass and the stellar mass assembly history (in the following denoted with a subscript ``t'' and ``s'', respectively).  

After examining the correlations between the various galactic properties (including $z_{\rm 50,s}$ and $z_{\rm mah,s}$) and the halo formation time ($z_{\rm 50,t}$, $z_{\rm mah,t}$),
 we have found that the best correlation is between $z_{\rm mah,s}$ and $z_{\rm mah,t}$ (with Pearson correlation coefficient $r=0.83$), followed by that between $z_{\rm 50,s}$ and $z_{\rm 50,t}$ ($r=0.54$), age and $z_{\rm 50,t}$ ($r=0.49$), and $z_{\rm 50,s}$ and $z_{\rm mah,t}$ ($r=0.36$).
Therefore, if the \citet{guo11} model is a good approximation to the real galaxy populations (see e.g., \citealt{lin13} for the agreement between the model prediction and observation for the stellar mass assembly history of brightest cluster galaxies at high redshift), with a suitable choice of formation time indicators, it is possible to infer the halo formation history from that of the central galaxies, for the low mass halos we consider here.
Furthermore,
if $z_{\rm 50,s}$ derived from VESPA is representative of the true value on average, our results presented in Section~\ref{sec:sfh} are on solid footing.

Ideally, we would like to calculate $z_{\rm mah,s}$ for the observed galaxy populations; however, we refrain from deriving it from the VESPA-based SFH here as the binning in lookback time for the public VESPA data is not optimal for this purpose.
Furthermore, it remains to be seen if VESPA-based SFH would result in a unbiased $z_{\rm mah}$.
In future work we plan to also investigate the use of other formation time indicators such as the strength of the 4000\,\AA\ break and the luminosity weighted mean age for the study of assembly bias.

In principle, the signature of assembly bias would be stronger if one uses the extrema of the distribution.  
Due to the temporal resolution of VESPA SFH, in our analysis, 
approximately we designate galaxies that have $z_{\rm 50,s}\gtrsim 1.8$ as early-forming, and those with $z_{\rm 50,s} \lesssim 2.1$ as late-forming (Section~\ref{sec:sfh}).
Should we have a higher resolution SFH, we could have examined the distribution of $z_{\rm 50,s}$ (or $z_{\rm mah,s}$), and only used the earliest- and latest-forming 20\% for the clustering and lensing measurements.  Such an analysis would require both much better quality spectra and much larger sample size (or much deeper imaging data than SDSS), however, and is therefore currently not yet feasible.

%%%%%%%%%%%%%%%%%%%%%%%%%%%%%%%%%%%%%%%%%%%
\subsection{Treatment of Galaxy and Halo Samples}
\label{sec:samples}
%%%%%%%%%%%%%%%%%%%%%%%%%%%%%%%%%%%%%%%%%%%

One potential concern regarding our way of constructing pure samples of central galaxies is the FOF removal of satellites.  While we believe this is a necessary operation, some of the galaxies thus removed may be those in the so-called backsplash halos, which may partially contribute to the signal of assembly bias at low mass scales (\citealt{wang07,dalal08}; see Section~\ref{sec:sfh}).
If true, it may complicate the interpretation of our results (in other words, the non-detection of assembly bias may be due in part to the removal of these galaxies).

Here we attempt to evaluate the effect of satellite removal on the magnitude of assembly bias using simulated data.
We repeat the procedure of Section~\ref{sec:sfh} and consider halo samples that have a mass distribution following the log-normal form.  The acceptable combinations of ($M_{\rm cen}, \sigma_{\log{M}}$) are again those that give $\chi^2 \le \chi^2_{\rm min}+2.3$ for a given observed galaxy sample. 
For a halo sample constructed from a given parameter set of ($M_{\rm cen}, \sigma_{\log{M}}$),
we remove halos that are located within $2r_{200c}$ from any of more massive halos in the whole
simulation box (that is, not restricted to those that satisfy the log-normal mass
distribution). 
For the remaining ``isolated'' halos, 
which may correspond crudely to our satellite-trimmed galaxy sample,
we apply the $z_{\rm form}>z_{\rm div}$ ($z_{\rm form}<z_{\rm div}$) criterion as before to further filter the halos, depending on whether the observed sample in question is early- or late-forming.
Finally, the theoretical expectation is obtained by averaging over the correlation functions from all acceptable models.
For the L0250 simulation used in Section~\ref{sec:sfh},
typically this procedure removes about 8\% of the halos, and reduces the ratio of the resulting correlation functions (early-to-late) by $\sim 10\%$ compared to that before the removal.  
Such a reduction would make theoretical predictions more compatible with our observations.
We find that the probability that the theory and the observed data are drawn from the same distribution is 
$p=0.055$ (with $z_{\rm form}=z_{\rm mah}$).
For the age-matching halo sample (Section~\ref{sec:ssfr}), 
about 30\% of the halos are removed, resulting in a larger reduction in the ratio of correlation functions, and
the probability for the model to be consistent with the observations is 
$p=0.0041$.

Admittedly this procedure is rather crude, but it should capture the essence of the effect of backsplash halos.  We see that indeed such halos may contribute partly to the assembly bias, which is consistent with the findings of \citet{wang09}.

Another potential concern is the form and width of the mass distribution of the real galaxy samples.  When matching an early-forming sample with a late-forming one, we only require their mean halo masses to be consistent (e.g., within $1\sigma$), but not the distribution in mass.
Our assumption that the halo mass distribution follows the log-normal form is informed by the age-matching model (Section~\ref{sec:sfh}), 
and with this form the centroid and width of the distribution may be roughly determined (as constrained by the observed $\Delta \Sigma$ profile). 
Although we have limited the mass range during the initial sample selection to be $\approx 0.5$ dex or smaller 
(based on the central stellar mass--halo mass relation, or the mass given in the Y07 catalog), 
we note these proxies are derived with the presence of satellites, and thus may bias the actual mass spread in an unpredictable way.  
A possible way forward is to construct (satellite-trimmed) central--halo relations with SFH or sSFR-selected samples (rather than those selected by stellar mass or luminosity thresholds, as commonly used), and use them to better constrain $(M_{\rm cen}, \sigma_{\log{M}})$ of the galaxy samples in question.

Finally, when we fit the NFW profile to the lensing measurements, we have assumed that the concentration is only a function of halo mass, and ignored any dependence of concentration on the halo formation time.  
This is mainly because the S/N of the galaxy-galaxy lensing profiles is insufficient to fit for concentration as well.
Such an assumption has the effect of potentially biasing the masses of our galaxy samples, in the sense that the mass of the early-forming (late-forming) sample would be overestimated (underestimated) when the mean concentration is used.  From the L0250 simulation, we have found that the concentration is $\sim 60\%$ higher in early-forming halos than in late-forming ones.  Adopting such a difference in our lensing measurements leads to roughly a 10\% reduction (increase) in halo mass for the early-forming (late-forming) galaxy sample.  Given that the difference in mass as presented in Section~\ref{sec:ours}
(i.e., when the mean concentration is used), as well as our mass measurement uncertainties, are both also at similar levels (10--20\%), 
and that observationally the difference in concentration of halos hosting red and blue central galaxies is  much smaller than the 60\% value used above  \citep{mandelbaum15},
we conclude that assuming concentration is only a function of halo mass does not have an appreciable effect in our analysis.

%%%%%%%%%%%%%%%%%%%%%%%%%%%%%%%%%%%%%%%%%%%
\subsection{Comment on Abundance Matching- and Satellite Kinematics-based Halo Mass Estimates}
\label{sec:yangmass}
%%%%%%%%%%%%%%%%%%%%%%%%%%%%%%%%%%%%%%%%%%%

In Sections~\ref{sec:prev} and \ref{sec:ours} we have used the halo mass estimates from Y07 and M11, which are based on an approach similar to abundance matching, and on satellite kinematics, respectively, to guide our initial sample selection.   
Although we have not carried out a systematic comparison of these estimates with weak lensing 
(see \citealt{mandelbaum15} for such an effort),
during our two-step procedure of constructing galaxy samples, we have  built three samples (selected only by color) as a by-product that can test the M11 relations.  
It appears that the red and blue central stellar mass--halo mass relations from M11 give rise to consistent halo masses, although these masses are $\sim 40\%$ 
higher 
than that indicated by lensing.  This is consistent with the finding of \citet{kravtsov14}.

As for the mass provided by Y07, which is obtained in a fashion similar to the abundance matching technique, our lensing measurements in Sections~\ref{sec:prev} and \ref{sec:ssfr} imply that the scatter in their mass estimates somehow correlates with basic physical properties of galaxies such as SFH or sSFR, at least at the low halo mass regime we study here.
Therefore, any study that assumes or requires the scatter in halo mass to be {\it random} should be cautious when adopting the mass estimates from this catalog.

%%%%%%%%%%%%%%%%%%%%%%%%%%%%%%%%%%%%%%%%%%%
\subsection{Implications of Our Results}
\label{sec:implications}
%%%%%%%%%%%%%%%%%%%%%%%%%%%%%%%%%%%%%%%%%%%

As we have demonstrated in Section~\ref{sec:sims}, together with numerous previous studies (e.g., \citealt{gao05,wechsler06,jing07,li08}), the halo assembly bias is a marked feature of the CDM model, especially at the low mass scales we study ($\sim 10^{12}\,h^{-1}M_\odot$).  How can this be reconciled with our lack of detection in galaxy populations?  It is possible that the baryonic processes of galaxy formation have rendered the signal small, 
the SFH derived by VESPA is too noisy so that the signal is washed out (for the case of Section~\ref{sec:sfh}),
the sSFR measurements from SDSS are too noisy to be a good indicator of the halo formation epoch (for the case of Section~\ref{sec:ssfr}),
or we have not yet found a galactic property that is closely linked to the halo formation history.

To check the first possibility, we again make use of the galaxy catalog from the \citet{guo11} model.  For central galaxies living in halos of mass $M_{200c}= (1-2)\times 10^{12}\,h^{-1}M_\odot$, we compute the mean of ratio $\xi_{\rm early}/\xi_{\rm late}$ over $5-20\,h^{-1}$Mpc, where the early- and late-forming samples are defined by several different conditions: (1) by the halo $z_{\rm mah,t}$, (2) by the stellar $z_{\rm mah,s}$, (3) by the stellar $z_{\rm 50,s}$, and (4) by the stellar age.  
For each of the formation time indicator, we designate the galaxies with value higher (lower) than the mode of the distribution to be early-forming (late-forming).
Comparing the ratio from cases (2)-(4) with that of case (1) would inform us whether the signature of assembly bias is erased or not with (this particular model of) galaxy formation.
It is found that the ratio for cases when the formation history is inferred from galactic properties is similar to, and not smaller than that derived purely from the dark matter halo assembly history, and thus it is probable that galaxy formation preserves assembly bias (c.f.~\citealt{wang13}).

It is certainly possible that the VESPA-based SFH is too noisy. 
Tests with mock galaxy spectra carried out in \citet{tojeiro09} indicate that for simple SFHs (exponential decay or dual-burst), the recovery of SFH is satisfactory.  The SFH of our central galaxies inevitably would be much more complicated than the simple cases tested above, and whether VESPA can reliably decipher the SFH needs to be checked with other tools such as STARLIGHT \citep{cidfernandes05}, FAST \citep{kriek09}, MAGPHYS \citep{dacunha08}, FIREFLY \citep{wilkinson15}.

In Section~\ref{sec:sims} we have noted that in the \citet{watson15} model, the sSFR of a central galaxy is assumed to have a one-to-one correspondence with the formation epoch of the host halo.  The assembly bias is thus {\it maximally} built-in in this model.
The fact that the probability for the model and our observations to be consistent is at most only at percent level (Section~\ref{sec:samples}) suggests
that either intrinsic scatters in the sSFR-formation epoch correspondence are much larger than assumed in the age-matching model (thus rendering the effect of assembly bias too small to be detectable), or the sSFR measurements from SDSS are not adequate for picking up the assembly bias signal, or a combination of both.

Finally, we entertain the possibility that we have yet to employ a new galactic property to better separate galactic systems into early- and late-forming ones, before we can unambiguously identify the assembly bias signature in the Universe.
 As discussed in Section~\ref{sec:formation}, the best candidate appears to be $z_{\rm mah,s}$, followed by $z_{\rm 50,s}$ and the mean age (see also the proxy discussed in \citealt{lim15}).  These in principle could be obtained from high quality spectra with VESPA, as well as the aforementioned codes,
 and will be subjects of our future investigation.
 Inspiration could also come from the analysis of \citet{miyatake15}, who have recently claimed a detection of strong assembly bias using galaxy clusters.  Their proxy for halo formation time is $\big< R_{\rm mem} \big>$, the mean projected separation of member galaxies from the cluster center.  In principle it is possible to adopt  a similar proxy for galaxy scale halos, although the number of satellites is much smaller, and the membership determination is less certain.

We conclude by noting that it is also imperative to better understand theoretically the origin of assembly bias across the mass spectrum of halos.  If the cause at low mass scales is the backsplash halos, then instead of identifying the best formation time indicator, one should look for ways to observationally distinguish such galactic systems, or to take such population fully into account when comparing with theoretical models.
Regarding the high mass end, since it is likely that the physical origin of assembly bias in high-mass halos is believed to be quite different than for low-mass halos \citep{dalal08}, the detection presented by \citet{miyatake15} is not necessarily inconsistent with the non-detection reported here.  
It is clear that the assembly bias phenomenon is far richer and complicated than expected, and warrants further investigations, such as exploring better proxies for halo formation time, and the scale dependence of assembly bias (Section~\ref{sec:magbias}).

%%%%%%%%%%%%%%%%%%%%%%%%%%%%%%%%%%%%%%%%%%%%%%%%%%%%%%%%%%%%%%%%%%%%%%
\section{Summary}
\label{sec:conclusion}
%%%%%%%%%%%%%%%%%%%%%%%%%%%%%%%%%%%%%%%%%%%%%%%%%%%%%%%%%%%%%%%%%%%%%%

As the assembly bias is a robust prediction of the CDM theory of structure formation, establishing it observationally would not only further vindicate this extremely successful theory, but also shed light on the baryonic physics of galaxy formation.
In this exploratory study of detection of assembly bias, we have shown (with the aid of weak
gravitational lensing) that some previous claims of detection may be simply due to  differences in halo mass of the galaxy samples, rather than a real manifestation of assembly bias (Section~\ref{sec:prev}).  We have then investigated a couple of ways of constructing galaxy samples with similar mean halo masses, thus facilitating a direct search for assembly bias in the real data (Section~\ref{sec:ours}).
We focus on the halo mass scale of $\sim 10^{12}\,h^{-1}M_\odot$, where assembly bias is expected to be large, and construct early- and late-forming halo samples by making use of a central galaxy catalog (Y07), under the assumption that the SFH of central galaxies reflects the formation history of underlying halos (which is well supported by a state-of-the-art galaxy formation model, Section~\ref{sec:formation}). 
Satellite galaxies living in massive halos that are misidentified as central galaxies in low mass halos we target would bias the mass estimates of the samples, and thus need to be removed.  We employ a FOF algorithm to achieve this.
Working with satellite-free samples, we consider two ways of inferring the formation epoch of the
central galaxies, namely the resolved SFH from VESPA (Section~\ref{sec:sfh}) and the current sSFR
(Section~\ref{sec:ssfr}).  

In both cases, after making sure the mean halo masses of the early- and late-forming samples are comparable with each other from weak lensing, we compare the relative large-scale bias of the samples with predictions from numerical simulations, 
finding the probability of the model and data to be consistent is very low.
We attribute this inconsistency to the possibilities that the formation epoch indicators we use are too noisy as derived from current data, or they do not correlate well with the actual halo formation history.
Although observational evidence for assembly bias remains elusive, we suggest a few indicators that should perform better for the distinction between early- and late-forming halos, which could be obtained with high signal-to-noise spectra of central galaxies (Section~\ref{sec:implications}).

%%%%%%%%%%%%%%%%%%%%%%%%%%%%%%%%%%%%%%%%%%%%%%%%%%%%%%%%%%%%
\section*{Acknowledgments}
%%%%%%%%%%%%%%%%%%%%%%%%%%%%%%%%%%%%%%%%%%%%%%%%%%%%%%%%%%%%%

YTL dedicates this work to the loving memory of his mother, Ms.~Chun-Chih Kang.
We are grateful to 
Keiichi Umetsu, David Spergel, Jim Gunn, Cheng Li, Rita Tojeiro, Zheng Zheng, Frank van den Bosch, Masamune Oguri, Uros Seljak, Lihwai Lin,  Chung-Pei Ma, and Ying Zu for helpful comments, 
to Gerard Lemson for help with the Millennium Simulation database,
to Surhud More for providing the data used in Fig.~\ref{fig:surhud} in electronic form, and to Bau-Ching Hsieh for help with parallel computation. 
We thank the anonymous referee for a constructive report.
YTL thanks I.H.~for constant encouragement and support.
YTL acknowledges support from the Ministry of Science and Technology grants MOST 102-2112-M-001-001-MY3 and MOST 104-2112-M-001-047.  
RM is supported by the Department of Energy Early Career Award Program and by a Sloan Fellowship.
This work was completed in part using the computing resources provided by the University of Chicago Research Computing Center. BD and AK were supported by the NASA ATP grant
NNH12ZDA001N and by the Kavli Institute for Cosmological Physics (KICP) at the
University of Chicago through grants NSF PHY-0551142 and PHY-1125897.

Funding for the SDSS and SDSS-II has been provided by the Alfred P.~Sloan Foundation, the Participating Institutions, the National Science Foundation, the U.S.~Department of Energy, the National Aeronautics and Space Administration, the Japanese Monbukagakusho, the Max Planck Society, and the Higher Education Funding Council for England. The SDSS is managed by the Astrophysical Research Consortium for the Participating Institutions.
The SDSS Web Site is http://www.sdss.org/.
The Millennium Simulation databases used in this paper were constructed as part of the activities of the German Astrophysical Virtual Observatory (GAVO).

%%%%%%%%%%%%%%%%%%%%%%%%%%%%%%%%%%%%%%%%%%%%%%

\end{document}